\documentclass{article}
\usepackage[dvips]{graphicx,color}
\usepackage{amsmath,amssymb}  
\newcommand{\bce}{\begin{center}}
\newcommand{\ece}{\end{center}}
\newcommand{\bea}{\begin{eqnarray}}
\newcommand{\eea}{\end{eqnarray}}
\newcommand{\be}{\begin{equation}}
\newcommand{\ee}{\end{equation}}
\newcommand{\bd}{\begin{displaymath}}
\newcommand{\ed}{\end{displaymath}}
\newcommand{\bit}{\begin{itemize}}
\newcommand{\eit}{\end{itemize}}
\newcommand {\ben}{\begin{enumerate}}
\newcommand{\een}{\end{enumerate}}
\newcommand{\bdes}{\begin{description}}
\newcommand{\edes}{\end{description}}

\newcommand{\E}{\> = \>}
\newcommand{\EA}{&=&}
\newcommand{\EQ}{\> \equiv \>}

\newcommand{\non}{\nonumber\\}

\newcommand{\Def}{\> := \>}
\newcommand{\deF}{\> =: \>}
\newcommand{\bfl}{\begin{flushleft}}
\newcommand{\efl}{\end{flushleft}}

\newcommand{\q}{\quad}
\newcommand{\Q}{\qquad}
\newcommand{\lrp}{\left ( \, }    
\newcommand{\rrp}{\, \right ) }   
\newcommand{\lsp}{\left [ \, }    
\newcommand{\rsp}{\, \right ] }   
\newcommand{\Si}{{\rm Si}}
\newcommand{\Ci}{{\rm Ci}}
\newcommand{\yto}{\> \stackrel{y \to \infty}{\longrightarrow}\>}
\newcommand{\hto}{\stackrel{h \to 0}{\longrightarrow}}
\newcommand{\leftdelta}{\stackrel{\leftarrow}{\delta}\!\!}
\newcommand{\rightdelta}{\stackrel{\rightarrow}{\delta}\!\!}

\newcounter{abb}

\begin{document}

\setcounter{page}{1}
\setcounter{section}{0}
\setcounter{footnote}{0}
\setcounter{abb}{0}
\setlength{\fboxrule}{0.7mm}

\renewcommand{\baselinestretch}{1.1}
\renewcommand{\thesection}{\arabic{section}}
\renewcommand{\thesubsection}{\thesection\arabic{subsection}}
\renewcommand{\theequation}{\thesection.\arabic{equation}}
\renewcommand{\thefootnote}{\arabic{footnote}}

\thispagestyle{empty}

%

\vspace{2cm}

\bce

{\Large On the Numerical Evaluation of a Class of Oscillatory Integrals}

\vspace{0.2cm}
{\Large in Worldline Variational Calculations}
\vspace{0.3cm}

\vspace{1.2cm}
R. Rosenfelder 

\vspace{0.6cm}
Particle Theory Group, Paul Scherrer Institute, CH-5232 Villigen PSI, 
Switzerland 
\ece

\vspace{2.5cm}

\begin{abstract}
\noindent
Filon-Simpson quadrature rules are derived for integrals of the type\\
\hspace*{2cm}$\int_a^b dx \, f(x) \> \sin(xy)/(xy) $ \hspace*{1cm} and  
\hspace*{1cm} $\int_a^b dx \, f(x) \> 4 \sin^2(xy/2)/(xy)^2 $ \\
which are needed in applications of the worldline variational approach 
to Quantum Field Theory. These new integration rules reduce to the standard Simpson
rule for $ y = 0 $ and are exact for $y \to \infty$ when $a = 0$ and $f(0) \ne 0 $.
The subleading term in the asymptotic expansion is also reproduced more and more
precisely when the number of integration points is increased.
Tests show that the numerical results are indeed stable over a wide range
of $y$-values whereas usual Gauss-Legendre quadrature rules are more precise
at low $y$ but fail completely for large values of $y$. The associated
Filon-Simpson weights are given in terms of sine and cosine integrals
and have to be evaluated for each value of $y$. A Fortran program to calculate 
them in a fast and accurate manner is available. A detailed omparison is made with the
double exponential method for oscillatory integrals due to Ooura and Mori.

\end{abstract}

\newpage
\setcounter{equation}{0}

\section{Introduction}
\label{sec: intro}
\setcounter{equation}{0}
Many problems in theoretical and computational physics require numerical 
integration over rapidly oscillating functions which usually is considered a 
difficult, if not hopeless
problem. However, as Iserles has pointed out  in Ref. \cite{Iser1} 
this must not be the case: If the quadrature rule
for an integral of the type
\be
\int_0^h dx \> f(x) \, e^{i x y} \> \simeq \> \sum_{i=0}^N w_i f(x_i)
\label{Fourier}
\ee
includes the endpoints $x_0 = 0, x_N = h $ then the error can be reduced to   
${\cal O} \left ( h^{N+2}/y^2 \right )$
for $ y \to \infty$. In Iserles' words: {\it As long as right methods are used, 
quadrature of highly-oscillatory integrals is very accurate and affordable !}
This result rules out the usual
quadrature schemes of Gaussian type but is realized in the classic Filon 
integration rules \cite{Filon} which are covered in many standard books
(see, e.g. chapter 2.10 in Ref. \cite{DaRa} or Eqs. 25.4.47 -
25.4.57 in Ref. \cite{Handbook}).

\vspace{0.5cm}
The present note is not concerned with the numerical evaluation of the Fourier
integral (\ref{Fourier}) which is well treated in numerous works but with 
oscillatory integrals whose integrand has a {\it removable singularity} at the
origin, viz.
\be
I_1[f](a,b,y) \Def \int_a^b dx \> f(x) \, \frac{\sin(x y)}{x y} 
\label{I1}
\ee
and
\be
I_2[f](a,b,y) \Def \int_a^b dx \> f(x) \> 4 \, \frac{\sin^2(x y/2)}{x^2 y^2} \> .
\label{I2}
\ee
The corresponding weight functions have been normalized in such a way that
they become unity for vanishing external (frequency) parameter $y$. In 
Eq. (\ref{I2}) one could rescale $ y \to 2 y$ to obtain a simpler 
oscillating weight function. However, due to $ \sin^2 t = ( 1 - \cos( 2t )/2 $ 
it would oscillate with twice the frequency and therefore
we prefer to define $I_2$ in the above form. This is also how it appears  
in the applications we will discuss below.
Note the relation
\be
I_1[f](a,b,y) \E \frac{1}{2 y} \frac{\partial}{\partial y} \, \left [ \, y^2 
I_2[f](a,b,y) \, \right ]
\> .
\label{I1 from I2}
\ee 

\vspace{0.2cm}
\noindent
As frequently is the case the necessity to evaluate  numerically integrals of 
the type (\ref{I1}, \ref{I2}) arose from a concrete theoretical problem. 
Here it was the polaron
problem in solid state physics \cite{polaron} and its extension, 
the worldline variational approach to relativistic Quantum Field Theory 
\cite{WC1,WC2,WC3,WC4,WC56,WC7,BRS}. With a general quadratic trial action
the Feynman-Jensen variational principle gives rise  
to a pair of non-linear variational
equations for the ``profile function'' $A(E)$ and the ``pseudotime''
$\mu^2(\sigma)$. For example, when calculating the self-energy of a single 
scalar particle in $d$ space-time dimensions these equations are of the form
\bea
A(E) \EA 1 + \frac{8}{d}\, \frac{1}{E^2} \int_0^{\infty} d\sigma \> 
\frac{\delta V}{\delta \mu^2(\sigma)} \> 
\sin^2 \left ( \frac{E \sigma}{2} \right ) 
\label{var A(E)}\\
\mu^2(\sigma) \EA \frac{4}{\pi} \int_0^{\infty} dE \> \frac{1}{A(E)} \, 
\frac{\sin^2(E \sigma/2)}{E^2} \> .
\label{amu2}
\eea
Here $ \> V[\mu^2] \> $ is the interaction term averaged over the trial action.
It is  a functional of the pseudotime and specific for the field
theory under consideration. For small proper times $\sigma$ 
its functional derivative (or ``force'') has the behaviour
\be
\frac{\delta V}{\delta \mu^2(\sigma)} \> \stackrel{\sigma \to 0}{\longrightarrow}
\> \frac{{\rm constant}}{\sigma^{d - 2 + r}}
\ee
where $ r = 0 $ for super-renormalizable theories and $r = 1 $ for renormalizable
ones like Quantum Electro-Dynamics (QED). Thus for the 3-dimensional 
polaron problem
and for super-renormalizable theories in 4 dimensions no divergences are 
encountered in the variational equations but QED$_4$ needs extra regularization, 
e.g. a cut-off at high momentum/small proper time.
In any case, at least the relation (\ref{amu2}) between profile function and 
pseudotime requires evaluation of integrals of the type (\ref{I2}) and 
frequently the proper-time derivative 
\be
\frac{d \mu^2(\sigma)}{d \sigma} \E \frac{2}{\pi} \, \int_0^{\infty} dE \> 
\frac{1}{A(E)} \, \frac{\sin (E \sigma)}{E} \> ,
\label{dotamu}
\ee
i.e. integrals of the type (\ref{I1}) are also needed. 

Of course, Eqs. (\ref{var A(E)}, \ref{amu2}, \ref{dotamu}) involve 
infinite upper limits
but the large-$E, \sigma$ limit of profile function and pseudotime may be 
obtained analytically so that we can restrict ourselves to finite 
integration limits and add the analytically calculated asymptotic contribution
to the numerical result. Numerical evaluation of Fourier integrals with 
infinite upper limits is much more demanding; one strategy
(employed in the adaptive NAG routine D01ASF \cite{NAG} based on 
Ref. \cite{QUADPACK}) requires a delicate extrapolation procedure and thus
seems to be not suitable for a numerical solution of the variational 
equations. 

Previously we have solved these equations 
on a grid of Gaussian points by iteration (for details see refs. 
\cite{WC2}) using Gauss-Legendre quadrature for the oscillatory
integrals. Although sufficient for many purposes problems of numerical 
stability became more serious, in particular in  QED$_4$ 
when the momentum cut-off was increased too much. It should be mentioned
that in the 1-body sector the profile function can be eliminated alltogether 
giving rise to a non-linear
integro-differential equation for the pseudotime only which shows a striking
resemblance to the classical Abraham-Lorentz-Dirac equation \cite{VALE}.
This eliminates the need for numerical evaluation of oscillatory integrals
but requires solution of non-linear delay-type equations. 
In addition, at present this is restricted to the 1-body case and 
for the 2-body, bound-state case we had to apply the previous scheme based on
(several) profile functions and pseudotimes \cite{BRS,varbind}.
Hence there is a definite need to have a reliable, fast and stable method for 
evaluating integrals of the type (\ref{I1}, \ref{I2}).
 
In the following sections these integration rules are derived ``naively'' 
(that is without mathematical rigor and error estimates) and 
tested for simple cases. Of course, there is a rich mathematical literature 
on quadrature rules for various or general oscillatory integrals
(for recent developments see, e.g., refs. \cite{Iser2,Iser3,Iser4}) 
but here the emphasis is more on the practical implementation and availability.
In addition, the behaviour for large values of the parameter $y$ will be 
investigated in some detail.

\section{Filon-Simpson quadrature}
\setcounter{equation}{0}
The strategy for deriving a stable quadrature formula for oscillating integrals
of the type
\be
\int_a^b dx \> f(x) \> O_j(xy) \simeq \sum_{i=0}^N w_i^{(j)} f \left (
x_i^{(j)} \right ) \> , \> \>  
O_j(xy) \E \left \{ \begin{array}{r@{\quad:\quad}l}
                    \frac{\sin(xy)}{xy} & j = 1 \\
                    \frac{4 \sin^2(xy/2)}{x^2 y^2} & j = 2
                    \end{array} \right . 
\label{def Oj}
\ee
is well known \cite{Iser1}: choose $N$ points $x_i^{(j)}$ 
in the interval $ [a,b] $, two of them identical with the endpoints
and require that the integral over 
$x^k  O_j(xy) $ is exact. This gives a system of equations for the integration
points $x_i^{(j)}$ and weights $w_i^{(j)}$. 

\vspace{0.3cm}
\noindent
Here, for simplicity, we choose equidistant, $j$-independent points
\be
x_i^{(j)} \EQ x_i \E a + i h \> , \> \> i = 0 \ldots N \> , \> \> 
h \E \frac{b-a}{N}
\ee
and $N = 2 $. In other words: this will be a generalization of 
Simpson's time-honoured rule, to which it should reduce for $O_j(xy=0) = 1$. 
It is for
this reason that we may call it {\it Filon-Simpson quadrature}. The integration
points being fixed there are three monomials which can be integrated exactly
giving rise to three equations for the weights 
\be
\sum_{i=0}^2 w_i^{(j)} x_i^k \E 
\int_a^{b = a + 2 h} dx \> x^k \>  O_j(xy) 
\Def J_k^{(j)} \> , \> \> \> k = 0, 1, 2 \> \> .
\ee
It is easy to solve this linear system of equations with the result
\bea
w_0^{(j)} \EA \frac{1}{2 h^2} \, \left [ \, x_1 \, x_2  \, J_0^{(j)} - \left ( x_1 
+ x_2 \right ) \, J_1^{(j)} +  J_2^{(j)} \, \right ] 
\label{w0}\\
w_1^{(j)} \EA \frac{1}{2 h^2} \, \left [ \, - 2 x_0 \, x_2 \, J_0^{(j)} + 
4 \, x_1 \, J_1^{(j)} - 2  J_2^{(j)} \, \right ]
\label{w1}\\
w_2^{(j)}  \EA \frac{1}{2 h^2} \, \left [ \, x_0 \, x_1  \, J_0^{(j)} - \left ( x_0 
+ x_1 \right ) \, J_1^{(j)} +  J_2^{(j)} \, \right ] \> .
\label{w2}  
\eea
Note that the weights depend on the lower and upper integration limit 
as well as on the external parameter $y$:
\be
w_i^{(j)} \EQ w_i^{(j)}(a,b,y) \>, \> \> h \E \frac{b - a }{2} \> .
\label{depend of w}
\ee
The relation (\ref{I1 from I2}) translates into
\be
J_k^{(1)} \E \frac{1}{2 y} \frac{\partial}{\partial y} \, y^2 \, J_k^{(2)} 
\> , \> \> k = 0, 1, 2 \> 
\Rightarrow \> w_i^{(1)} \E \frac{1}{2 y} \frac{\partial}{\partial y} 
\, y^2 \, w_i^{(2)} \> , \> \> i = 0, 1, 2  
\ee
and the integrals may be written as
\be
J_k^{(j)} \E \frac{1}{y^{k+1}} \, \left [ \, F_k^{(j)}(by) - F_k^{(j)}(ay) \, 
\right ]
\label{Ikj as diff}
\ee
with
\be
F_k^{(j)}(z) \E \int_0^z dt \> t^k \, O_j(t) \> .
\ee
From the explicit form of the weight functions $O_k^{(j)}$ one finds
\bea
F_0^{(1)}(z) \EA \Si(z) \> , \> \> F_1^{(1)}(z) \E 1 - \cos z \> , \> \> 
F_2^{(1)}(z) \E \sin z - z \cos z 
\label{F^1}\\
F_0^{(2)}(z) \EA 2 \, \left [ \, \Si(z) - \frac{1-\cos z}{z} \, \right ] 
\> , \> \> 
F_1^{(2)}(z) \E 2 \, \left [ \, \gamma + \ln z - \Ci(z) \, \right ] 
\> , 
\label{log terms 1} \\
F_2^{(2)}(z) \EA 2 \, \left [ \, z - \sin z \, \right ] \> . 
\label{F^2} 
\eea
Here $ \gamma = 0.5772156640 \ldots $ is Euler's constant and $\Si(z), \Ci(z)$ the
sine and cosine integral, respectively, as defined in chapter 5 of Ref. 
\cite{Handbook}. Note that the integrals $J_k^{(j)}$ are odd 
under exchange $ a \leftrightarrow b $ as may be seen from their definition or
from Eq. (\ref{Ikj as diff}). Therefore we find
\be
w_i^{(j)}(b,a,y) \E - \, w_{2-i}^{(j)}(a,b,y) \> \> , \> \> i \E 0, 1, 2
\ee
which reflects the basic property of the integrals $I_j[f]$ when their limits
are exchanged.

\vspace{0.1cm}

As a check we now consider the limit $y \to 0 $. It is easily seen 
that the integrals $J_k^{(j)}$ and therefore the weights $w_i^{(j)}$ are
well-behaved in that limit because we have
\be
F_k^{(j)}(z) \E z^{k+1} \, \sum_{n=0}^{\infty} \, (-1)^n 
\frac{j}{(2n + j)!} \, \frac{z^{2n}}{2n + k + 1} \> .
\ee
Therefore 
\be
J_k^{(j)} \> \stackrel{y \to 0}{\longrightarrow} \> \frac{1}{k+1} \, \left [ 
b^{k+1} - a^{k+1} \right ] + {\cal O} \left ( y^2 \right ) \> .
\ee
Eqs. (\ref{w2} - \ref{w0}) and $ b = a + 2 h $ then immediately give the standard
Simpson weights
\be
w_0^{(j)}, \, w_1^{(j)}, \, w_2^{(j)}   \> \stackrel{y \to 0}{\longrightarrow} 
\> \frac{h}{3}, \,\frac{4h}{3},\,  \frac{h}{3} + {\cal O} \left ( y^2 \right ) 
\> , 
\ee
independent of the lower limit $a$ and the type $j$ of the integral as it should 
be.

More interesting is the limit $ y \to \infty $ for the case $a = 0$. First, the 
exact asymptotic behaviour of the integrals $I_j$ is easily obtained by 
substituting $ t = x y $ :
\be
I_j[f](a=0,b,y) \E \frac{1}{y} \, \int_0^{by} dt \> f \left ( \frac{t}{y} 
\right ) \, O_j(t)  \> \yto \> 
\frac{f(0)}{y} \, \int_0^{\infty} dt \> O_j(t)  \E \frac{f(0)}{y} \, 
\frac{j \pi}{2} \>, \> \> j = 1, 2 \> .
\label{In asy}
\ee
The asymptotic limit of the weights is found by
employing the expansions \cite{Handbook}
\be
\Si(z) \> \stackrel{z \to \infty}{\longrightarrow} \> \frac{\pi}{2} - 
\frac{\cos z}{z}  - \frac{\sin z}{z^2}+ \ldots \> , \hspace{1cm} 
\Ci(z) \> \stackrel{z \to \infty}{\longrightarrow} \> \frac{\sin z}{z}  + \ldots
\label{sinint asy}
\ee
which leads to
\bea
J_0^{(j)} & \yto& \frac{j}{y} \, \frac{\pi}{2} + 
{\cal O} \left ( \frac{1, \cos by}{y^2} \right ) \non
J_1^{(j)} & \yto & \frac{2j-2}{y^2} \ln by + {\cal O} \left ( \frac{1,\cos by}{y^2} 
\right )
\label{log terms 2} \\
J_2^{(j)} & \yto & {\cal O} \left ( \frac{1,\cos by}{y^2} \right ) \nonumber \> .
\eea
Therefore $J_0^{(j)}$ provides the leading asymptotic contribution and  
from Eqs. (\ref{w0} - \ref{w2}) we see that
for $x_0 = a = 0$ only $w_0^{(j)}$ is affected by it:
\bea
w_0^{(j)}  & \yto & \frac{j}{y} \, \frac{\pi}{2} + 
{\cal O} \left ( \frac{(2j-2) \ln by, 1, \cos by}{y^2} \right ) \\
w_1^{(j)}, \,  w_2^{(j)}  & \yto & 
{\cal O} \left ( \frac{(2j-2) \ln by, 1, \cos by}{y^2} \right ) \> .
\eea
Hence Filon-Simpson quadrature gives in that limit
\be
\sum_{i=0}^{2} w_i^{(j)} f(ih) \> \yto \> 
w_0^{(j)} f(0) \E \frac{j}{y} \frac{\pi}{2} \, f(0)
\ee
in perfect agreement with the leading asymptotic result (\ref{In asy}) !

Of course, the story is different if the function $f(x)$ vanishes at $ x = 0 $:
then the next-to-leading terms of the asymptotic expansion come into play. These
are studied in more detail in Appendix A1 and A2 where it is shown that for small 
enough increment $h$ the subleading terms are also reproduced approximately. 
For example, 
derivatives are replaced by finite differences as is expected from a quadrature 
rule which is based on values and not derivatives \cite{Iser4} of the function 
to be integrated. 

\section{Extended Filon-Simpson rule}
\setcounter{equation}{0}
We would like to decrease the error of the quadrature rule in a systematic way
without messing up the construction. For equidistant integration points this is, 
of course, very easy to achieve by dividing the integration interval in 
$ N/2 $ sub-intervals, $N$ being even, and applying the integration rule in each 
sub-interval. We then obtain the `extended' (or 'composite') Filon-Simpson
quadrature rule
\be
\boxed{
\int_a^b dx \> f(x) \, O_j(xy) \> \simeq \> \sum_{i=0}^{N} W_i^{(j)} \, 
f(a + i h) \>} \>  , \> \> h \E \frac{b-a}{N} \> , \> \> N \> {\rm even}
\ee
where (using the explicit nomenclature of Eq. (\ref{depend of w}))
\bea
W_0^{(j)}  \EA w_0^{(j)}(a,a+2h,y) \\
W_i^{(j)}  \EA w_1^{(j)}\left (a+(i-1)h,a+(i+1)h,y \right ) \hspace{1cm}
i = 1,3 \ldots (N-1) \\
W_i^{(j)}  \EA w_2^{(j)}\left (a+(i-2)h,a+ih,y \right ) 
+  w_0^{(j)} \left (a+ih,a+(i+2)h,y \right ) \hspace{0.5cm} i = 2,4 \ldots (N-2) \\
W_N^{(j)}  \EA w_2^{(j)}\left (a+(N-2)h,a+Nh,y \right ) \> .
\eea
Because of the complicated dependence of the weights $w_i^{(j)}$ on the 
lower and upper integration limit it is not possible to simplify 
the above expressions for $y \ne 0 $. Only for $ y = 0 $ we obtain the weights 
for the extended Simpson rule 
\be
W_i^{(j)} \> \stackrel{y \to 0}{\longrightarrow} \> \frac{h}{3} 
\left \{ \begin{array}{r@{\quad:\quad}l}
         1 & i = 0,N \\
         4 & i = 1,3 \ldots (N-1)\\
         2 & i = 2,4  \ldots (N-2) \> . \end{array} \right. 
\ee
The asymptotic limit $y \to \infty$ for $a = 0 $ is also unchanged from 
the the previous section: this is because only in the first interval from $a =0$
up to $2h$ the leading term $\pi/2$ from the asymptotic expansion 
(\ref{sinint asy}) of the sine integral survives whereas in the next and 
all other intervals it is cancelled due to difference which has to be taken 
in Eq. (\ref{Ikj as diff}). Thus in the extended Filon-Simpson quadrature rule 
asymptotic high oscillations are also treated correctly in leading order.
Appendix A3 shows that this also holds for the subleading term if the increment
$h$ is made small enough, i.e. if the number $N$ of integration points is increased.
This is a remarkable feature since logarithmic terms show up in the next-to-leading
order of the asymptotic expansion of $I_2[f]$ (see Appendix A1). However, 
Eq. \eqref{log terms 1} shows that by construction such terms are also present in the 
Filon-Simpson rule for $I_2[f]$.

\section{Numerical tests}
\setcounter{equation}{0}
Here we report test evaluations of the integrals
\be 
I_j \left [f_l = x^l e^{-x} \right ] \Bigl ( a=0,b=\infty,y \Bigr ) 
\E \int_0^{\infty} dx \> x^l \, e^{-x} \, O_j(xy)
\ee
for $ l = 0, 1 $ with the Filon-Simpson and other quadrature rules. The 
oscillatory weight functions $O_j(xy)$ have been defined in Eq. (\ref{def Oj}).
The exact values of the test integrals are (see e.g. p. 234 -- 235 in Ref. 
\cite{Dwight})
\bea
I_1[f_0] \EA \frac{\arctan(y)}{y} \> , \> \> I_1[f_1] \E \frac{1}{1 + y^2}
\label{exact1}\\
I_2[f_0] \EA  \frac{1}{y^2} \left [ 2 y \arctan(y) - \ln \left ( 1 + y^2 
\right )  \right ] \> , \>  \> I_2[f_1] \E  \frac{1}{y^2} \ln \left ( 1 + y^2 
\right )  
\label{exact2}
\eea
\refstepcounter{abb}
\begin{figure}[hbt]
\bce
\includegraphics[angle=0,scale=0.5]{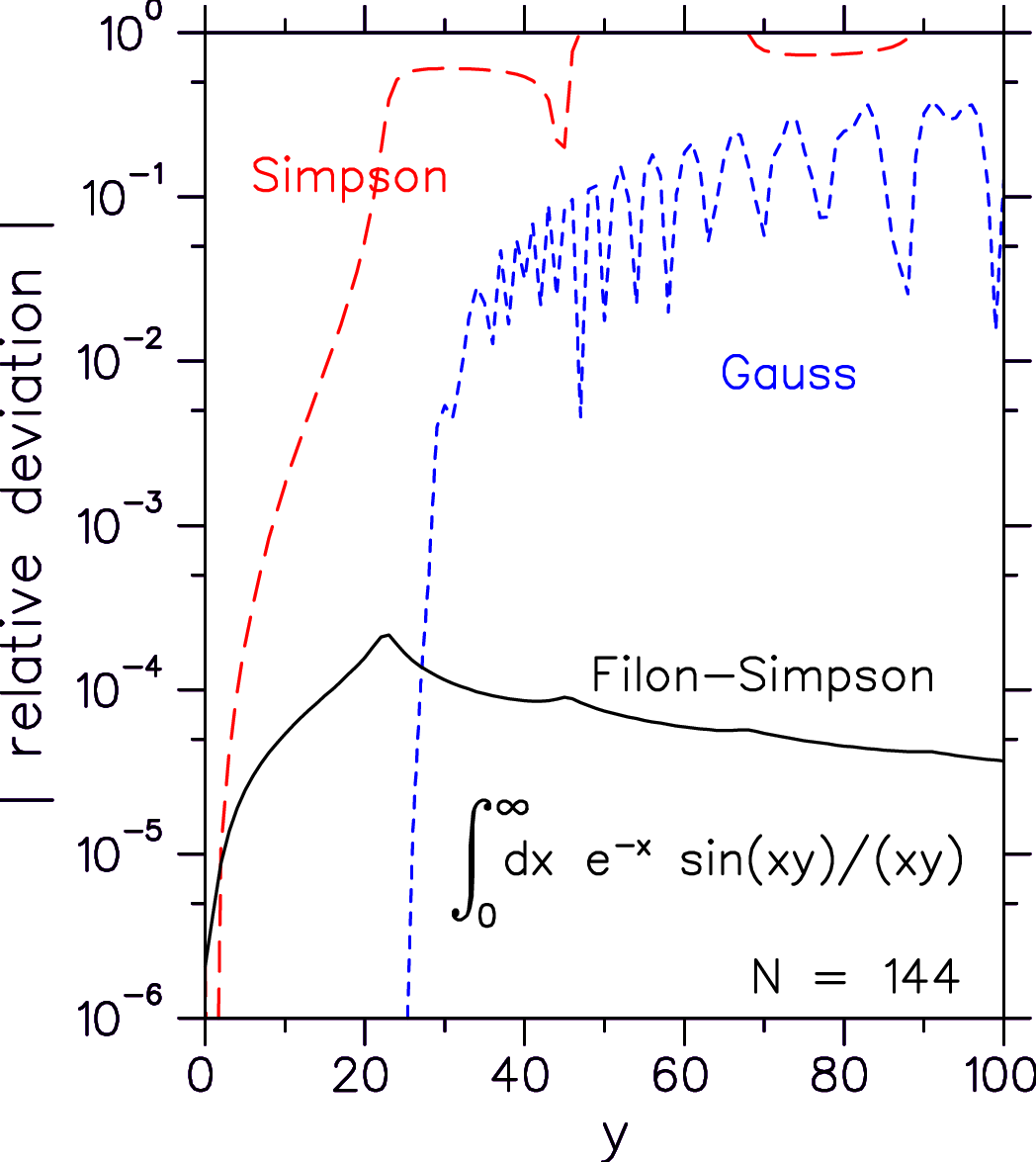}
\ece
\vspace{0.5cm}
\caption{Absolute value of the relative deviation between numerical and exact 
result for the oscillatory integral $I_1[f_0]$ 
as a function of the frequency parameter $y$.
The numerical result was obtained with different quadrature rules by
numerical integration up to $b = 20$ and adding the asymptotic contribution
(\ref{asy1}). The integrand is 
the function $f_0(x) = e^{-x}$ times the oscillatory weight function 
$O_1(xy) = \sin(xy)/(xy)$ and $144$ integration points have been used. 
}
\vspace{0.6cm}

\label{fig:I1_f0_144}
\end{figure}

\noindent
and fulfill the relation (\ref{I1 from I2}). As discussed 
in the Introduction our strategy is to perform the numerical integration 
up to $x = b$ and to add the asymptotic contribution
\be
\Delta_{jl} \Def \int_b^{\infty} dx \> x^l \, e^{-x} \, O_j(xy) \> .
\ee
Due to the specific form (\ref{def Oj}) of the weight functions $O_j$
it can be expressed in terms of the exponential integral \cite{Handbook}, viz.
\be
\Delta_{1l} \E \frac{b^l}{y} \, {\rm  Im} \, 
{\rm E}_{1-l} \left ( (1-iy)b \right ) \> ,
\> \> \Delta_{2l} \E \frac{2 b^{l-1}}{y^2} \, {\rm Re} \, 
\Bigl [ \, {\rm E}_{2-l}(b) 
-   {\rm E}_{2-l} \left ( (1-iy)b \right ) \, \Bigr ] \> .
\ee
The asymptotic expansion of these functions gives
\bea
\Delta_{1l} & \simeq & \frac{b^{l-1}}{1 + y^2} \, e^{-b} \, 
\left [ \, \cos by + 
\frac{\sin by}{y}  + {\cal O} \left ( \frac{1}{b} \right ) \, \right ] 
\label{asy1}\\
\Delta_{2l}  & \simeq & 2 \frac{b^{l-2} }{1 + y^2}  \, e^{-b} \, \left [ \, 
\frac{1 + y^2 + y \sin by - \cos by}{y^2}  + 
{\cal O} \left ( \frac{1}{b} \right ) \, \right ]
\label{asy2}
\eea
which is well-behaved even for $ y = 0 $.
In all numerical calculations we have chosen 
\be
b \E 20
\ee
so that in general only a very small fraction 
$ {\cal O}\left ( e^{-20} \right ) \simeq 5 \cdot 10^{-8} $ times powers of 
$b$ comes from the asymptotic region. We have checked that the next order
of the asymptotic expansion does not alter the outcome on the required accuracy 
level except at $y = 0$ where small changes are observed.

The Filon-Simpson weights are given in terms of sine and cosine integrals so that
a fast and precise routine for these special functions is mandatory. Fortunately
there are many convenient rational approximations available, e.g. in chapter 5
of Ref. \cite{Handbook} or routines from commercial libraries.
However, with little extra effort the Chebyshev 
expansions given in Table 23 of Luke's comprehensive book \cite{Luke}  
provide accurate values for these functions which exceed the double precision
arithmetic which we employ. Therefore we have used these expansions
in a Fortran subroutine making the whole program for calculating 
Filon-Simpson weights self-contained and portable \footnote{A copy of the program 
together with a sample run is available on request.}.

\begin{figure}[hbt]
\bce
\includegraphics[angle=0,scale=0.5]{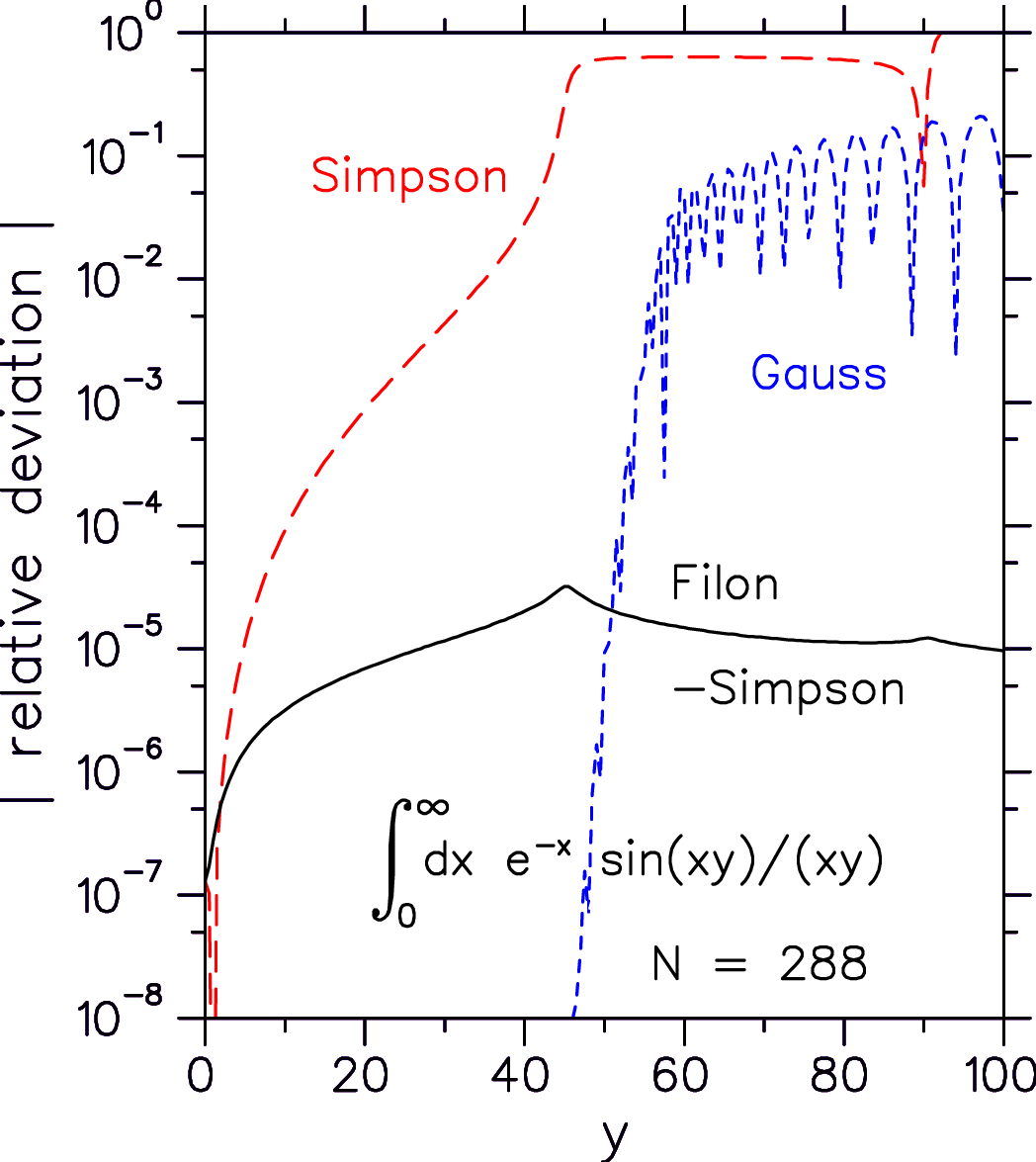}
\ece
\caption{As in Fig. \ref{fig:I1_f0_144} but with $288$ integration points. Note
the extended scale for the relative deviations.}
\label{fig:I1_f0_288}
\end{figure}

Fig.  \ref{fig:I1_f0_144} shows the relative deviation of the 
numerically calculated integral
$I_1[f_0]$ plus asymptotic contribution from the exact result (\ref{exact1})
as a function of the external variable $y$. $144$ integration points have 
been used for both the Filon-Simpson and Simpson rule and $2 \times 72$ points
for the Gauss-Legendre quadrature (i.e. subdivision of the whole interval in 
2 parts and application of a 72-point Gauss-Legendre rule in each part). 
In Fig.  \ref{fig:I1_f0_288} the
number of integration points is increased to $288$ and $4 \times 72$ respectively.
It is seen that for small $y$ the Gauss-Legendre rule is superior but starts 
to deteriorate when the number of integration points is insufficient for
the rapid oscillations of the integrand. 

By increasing the number of
integration points the onset of failure can be extended (from $ y \simeq 25$
in Fig.  \ref{fig:I1_f0_144} to $ y \simeq 50$ in Fig.  \ref{fig:I1_f0_288}) but
not avoided~\footnote{A rule-of-thumb is that one needs at least one
Gaussian integration point on each oscillation. Indeed, assuming that values
up to $x = b$ contribute significantly to the integrals we would have
$y_{\rm max} \simeq \pi N/b \simeq 0.15 N$ in qualitative agreement with Figs. 
\ref{fig:I1_f0_144}, \ref{fig:I1_f0_288}. }. 
The ordinary Simpson rule is even less capable to
deal with such type of integrals. In contrast, the new Filon-Simpson quadrature
rule gives stable results for all $y$-values considered and its relative
deviation from the exact result is a smooth function of $y$ {\it decreasing}
for very large $y$. Of course, the calculation of the weights has to be 
redone for each value of $y$ and is more involved than for the standard,
simple integration rules. However, in terms of CPU-time this is 
still a neglible expense (200 $y$-values in Fig.~\ref{fig:I1_f0_288} took 
about 2.5 seconds on a 600 MHz Alpha workstation).

\refstepcounter{abb}
\vspace{1cm}

\begin{figure}[hbt]
\bce
\includegraphics[angle=0,scale=0.45]{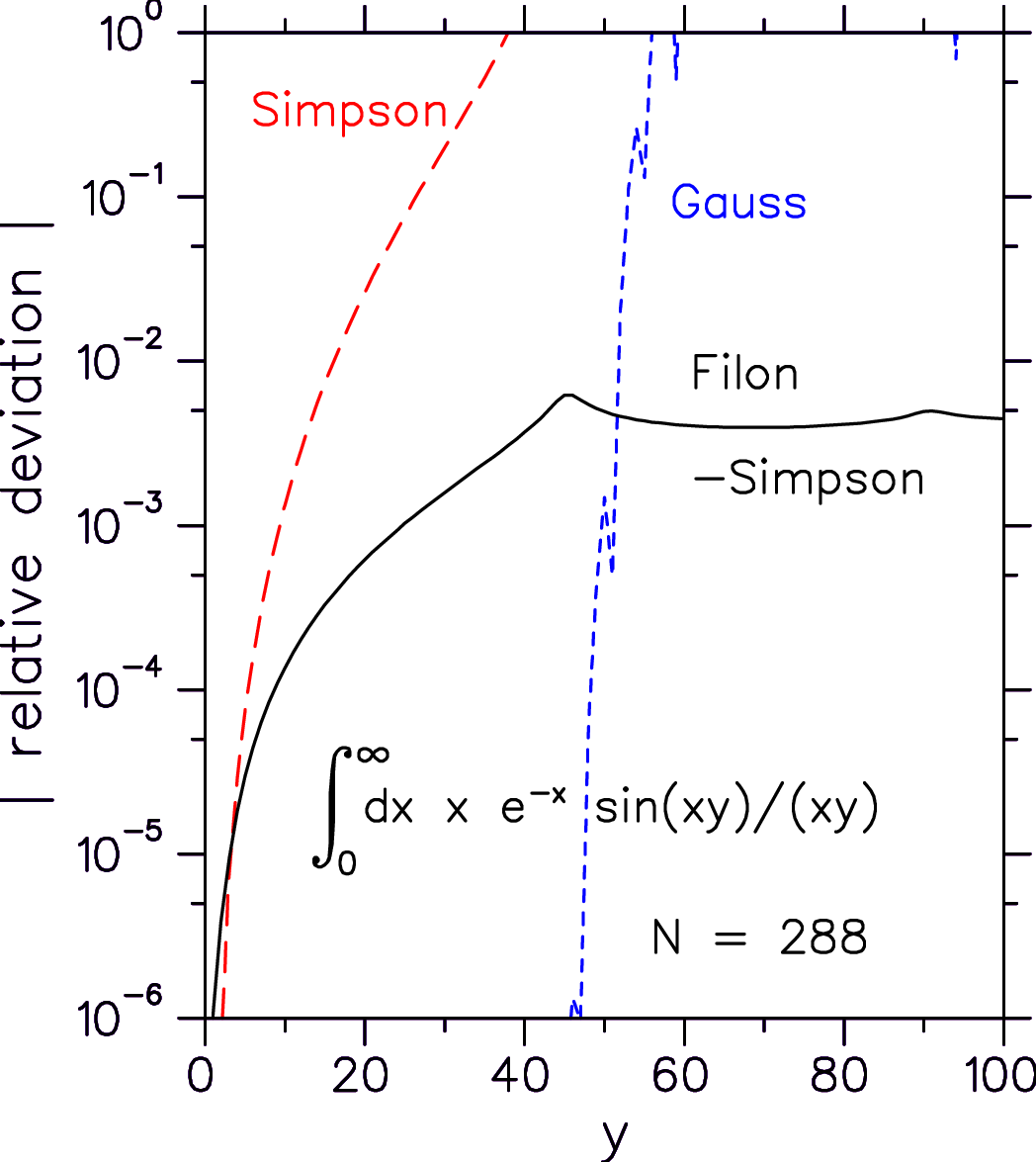}
\ece
\vspace{-0.3cm}

\caption{Same as in Fig. \ref{fig:I1_f0_144} but for
the test function $f_1(x) = x e^{-x}$ and $288$ integration points.
}
\label{fig:I1_f1_288}
\end{figure}

Fig. \ref{fig:I1_f1_288} depicts the result for the test function $f_1$ which has 
an additional $x$-power in the integrand so that the relative accuracy
which can be achieved with a fixed number of integration points is worse than
in the previous case. In addition now $f_1(0) = 0 $ so that the leading
asymptotic term vanishes but Appendix A3 demonstrates that the subleading term
is also reproduced for sufficiently large number of integration points. Therefore
Filon-Simpson integration still does far better than 
ordinary Simpson or Gauss-Legendre rules.

\begin{figure}[htb]
\bce
\includegraphics[angle=0,scale=0.45]{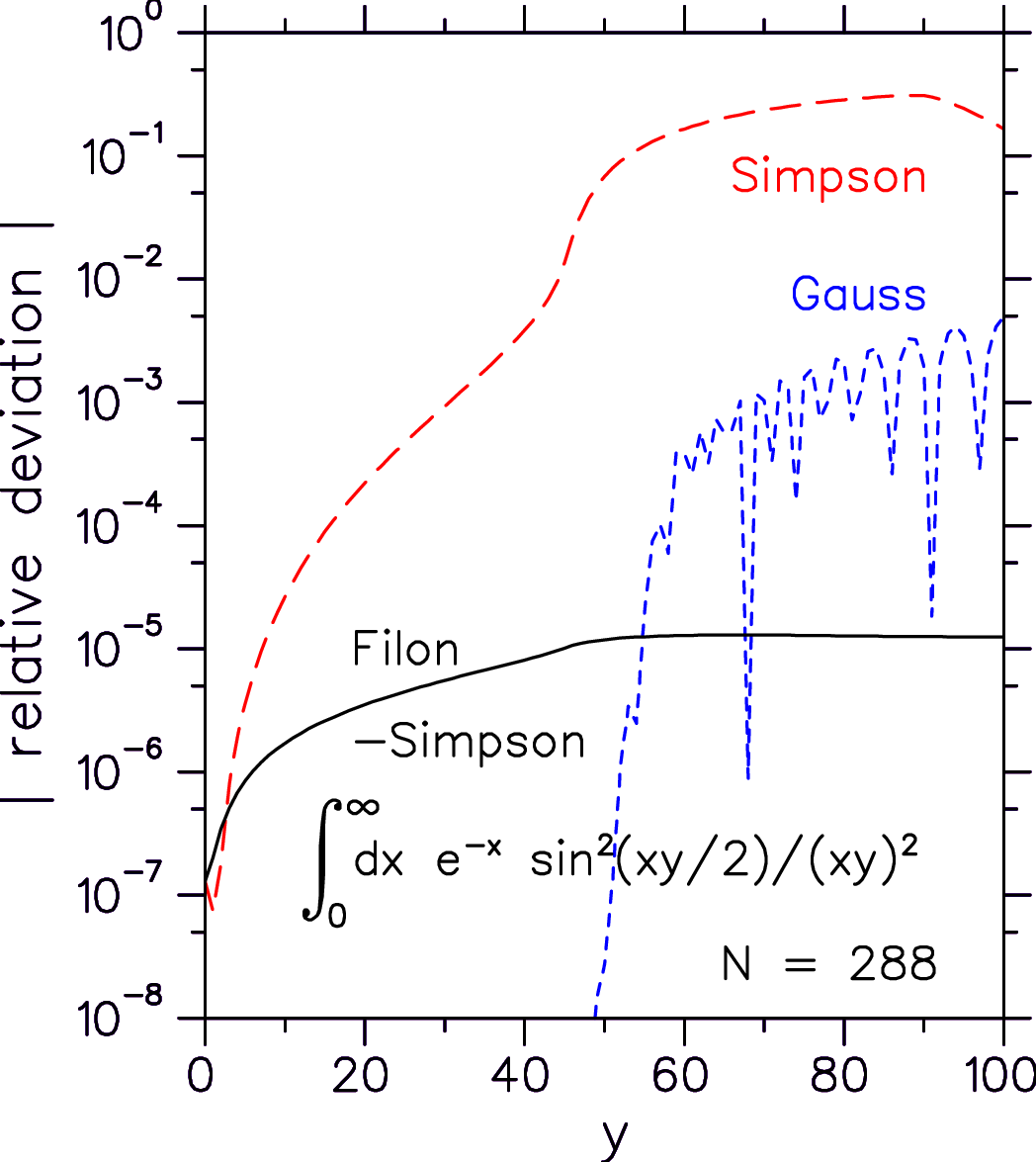}
\ece
\vspace*{-0.8cm}
\caption{Same as in Fig. \ref{fig:I1_f0_288} but with 
the oscillatory weight function $O_2(xy)$ and the asymptotic contribution 
(\ref{asy2}).}
\label{fig:I2_f0_288}
\end{figure}

\begin{figure}[hbt]
\bce
\includegraphics[angle=0,scale=0.4]{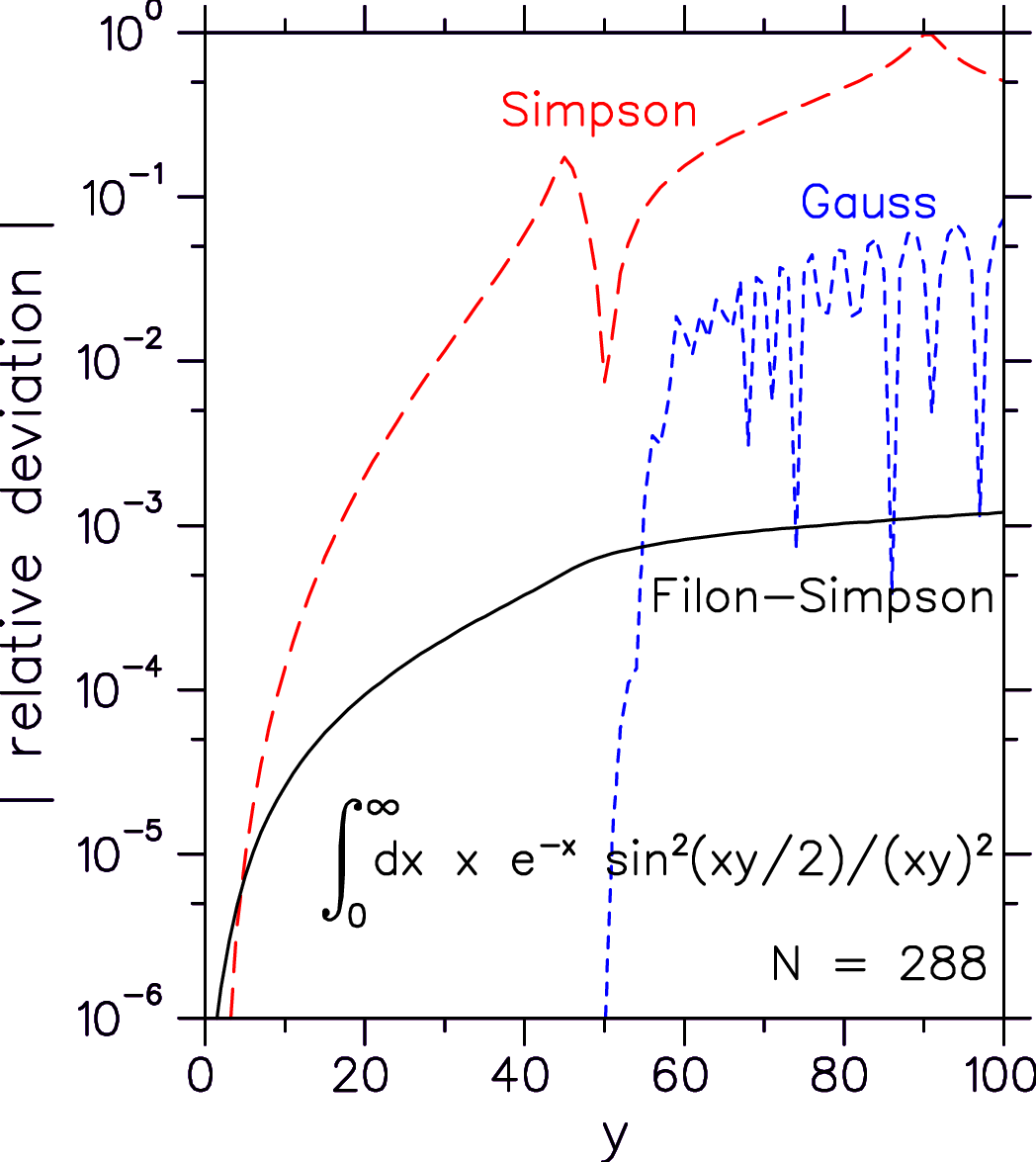}
\ece
\caption{Same as in Fig. \ref{fig:I1_f1_288} but with 
the oscillatory weight function $O_2(xy)$ and the asymptotic contribution 
(\ref{asy2}).} 
\label{fig:I2_f1_288}
\end{figure}

The corresponding results for the integral $I_2[f_l]$, i.e. with the
weight function $ 4 \sin^2(xy/2)/(xy)^2 $ are shown in Figs. \ref{fig:I2_f0_288}
and \ref{fig:I2_f1_288} and confirm the experience gained with $I_1$.

\refstepcounter{abb}

\refstepcounter{abb}

\refstepcounter{abb}

\noindent
Finally, we have investigated whether it is advantagous to use the relation 
\be
I_2[f](a,b,y) \E \frac{2}{y^2} \, \int_0^y dy' \> y' \, I_1[f](a,b,y')
\label{relation}
\ee
which is obtained from Eq. (\ref{I1 from I2}) by integration 
(the integration constant is zero since the integrals $I_j$ are finite 
at $ y = 0 $). Here one doesn't have to integrate over an oscillating 
function and the asymptotic limit (\ref{In asy}) is also correctly obtained.
In the worldline application this would amount to first evaluate 
$ d \mu^2(\sigma)/d\sigma$ from Eq. (\ref{dotamu}) and then integrate it
step by step via a trapezoidal or Simpson rule to obtain $\mu^2(\sigma)$.

\begin{figure}[htb]
\bce
\includegraphics[angle=0,scale=0.4]{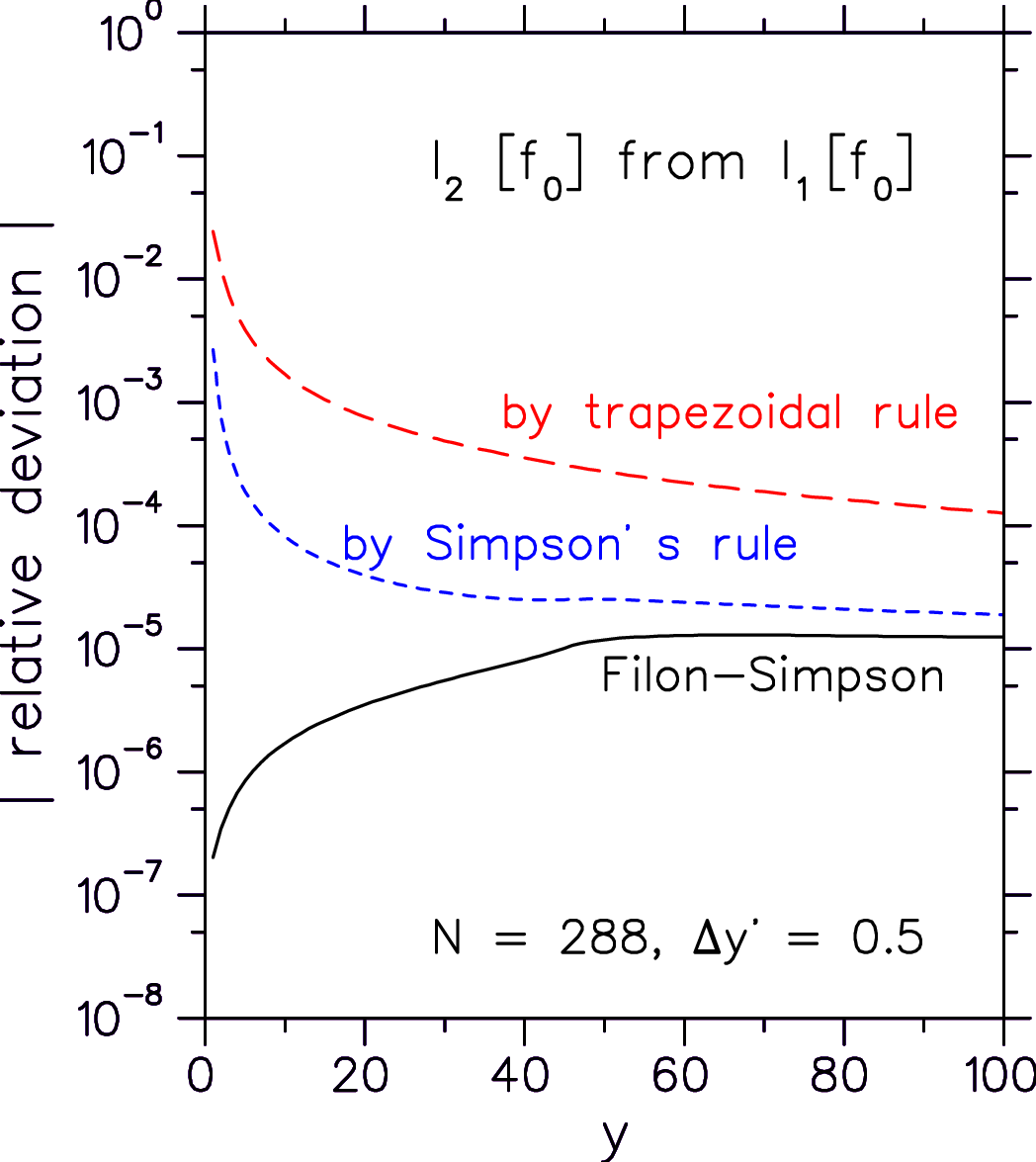}
\ece
\caption{Absolute value of the relative deviation between numerical and exact 
result if the
relation (\ref{relation}) is used to obtain the integral $I_2[f_0]$ from 
$I_1[f_0]$. The latter was calculated by the Filon-Simpson routine with 288 points
$+$ asymptotic contribution as function of $y'$ in steps of $\Delta y' = 0.5$
and then numerically integrated 
over $y'$ using either the trapezoidal rule or Simpson's rule. For comparison the 
result from the direct evaluation of $I_2[f_0]$ by the Filon-Simpson rule with
288 integration points $+$ asymptotic contribution (as in 
Fig. \ref{fig:I2_f0_288}) is also shown. 
}
\label{fig:relation_288_05}
\vspace*{1cm}
\end{figure}

Fig. \ref{fig:relation_288_05} shows the comparison with the direct Filon-Simpson
integration. While a lot of accuracy is lost at small $y$ with this procedure 
reasonable accuracy can be achieved at larger values of the frequency parameter 
$y$. However, high accuracy requires a precise and smooth input $I_1[f](a,b,y')$ 
together with a fine mesh of $y'$-values. Considering how fast and easy the 
Filon-Simpson weights can be generated this procedure does not offer real 
advantages and is not recommended.

\refstepcounter{abb}
\clearpage
%
%
\section{Comparison with the Double Exponential Method}
\setcounter{equation}{0}
The double exponential method of Takashi and Mori \cite{TaMori} is based
on the Euler-Maclaurin summation formula~\footnote{See Eq. 23.1.30 in Ref. \cite{Handbook}. 
Here $ h = (b - a)/N $, 
$B_{2k}$ are the Bernoulli numbers and $f$ is supposed to have $ 2m $ continous derivatives in $ [a,b]$ .} 
(or equivalently  the trapezoidal rule) 
\be
\int_a^b  dx  f(x)  \E h \sum_{k=0}^N f(a + k h) - \frac{h}{2} 
\left [ f(a) + f(b) \right ] - \sum_{k=1}^{m-1}\frac{h^{2k} B_{2k}}{(2k)!} \left [  
f^{(2k-1)}(b) -  f^{(2k-1)}(a)  \right ]  + R_m 
\ee
with
\be 
R_m \E - \frac{h^{2m+1} 
B_{2m}}{(2m)!} \, \sum_{k=0}^{N-1} f^{(2m)}(a + k h + \theta h) \> ,  \qquad 0 < \theta < 1
\ee
and the following observation:
When $f(x)$ and all its derivatives vanish at the endpoints $a$ and $b$ 
then the error of the trapezoidal approximation to the integral is given by $ R_m $ only 
and for an analytic function ($ m \to \infty $)
it goes to zero more rapidly than any power of $h$. Indeed if $ f(x) $ is
analytic in a strip $ \> |{\rm Im} \, t | < d \>  $ it has be shown \cite{Mori} that
\be
R_{\infty}  \> \sim \> e^{-{\rm const} \,  d/h}  \quad {\rm and} \quad \sim \> e^{-{\rm const'} \, N/\log N}  \> .
\ee
This property can be achieved by a special transformation  $ x = g(t) $ so that
\bea
\int_a^b dx \> f(x) \EA \int_{-\infty}^{+\infty} dt \> g'(t) \, f(g(t)) \non
\EA h  \sum_{k=-\infty}^{+\infty} w_k \, f(x_k)  + R(h) \quad
{\rm with}\hspace{0.2cm} x_k = g(k h) \hspace{0.2cm} {\rm and} 
\hspace{0.2cm} w_k = g'(k h) \> .
\label{double exp}
\eea
In the second line the trapezoidal rule for the infinite integral is used. 
The transformation proposed by Takashi and Mori \footnote{They take $ \lambda = \pi/2 $ as optimal 
but experimentation shows that $ \lambda = 1 $ is equally good.
Similar transformations exist for infinite and half-infinite intervals.} is
\be
x \E \frac{b + a}{2} + \frac{b-a}{2} \tanh \left [ \, \lambda \,  \sinh t \, \right ]  \>, \hspace{1cm} 
t \in [-\infty,+\infty] \> , \> \> \> \lambda > 0 
\ee
from which the alternative name "tanh-sinh integration rule" is derived (a short introduction is provided by Ref. \cite{BBBZ}, 
an overview is given in Ref. \cite{Mori-discov}.)
The infinite sum in 
Eq. (\ref{double exp}) may be truncated  without problems since for $ |k| \to \infty $ one has rapid,
``double exponential'' convergence
\be
w_k \> \longrightarrow \>  \lambda \,  ( b - a ) \,\exp \left ( \> - \lambda \, e^{|k| \, h} \> 
\right ) \> .
\ee 
\vspace*{0.2cm}

Ooura and Mori \cite{OoMori} (OM) have extended this scheme to oscillatory integrals. 
Here we describe the method for our integrals 
\be
I_j \E \int_0^{\infty} dx \> f(x) \,  O_j(xy) \quad {\rm with} \quad O_j(xy) \E \lrp \frac{\sin(xy/j)}{xy/j} \rrp^j \> , 
\quad j = 1, 2
\ee
in which $ f(x) $ is a non-oscillatory function.
Making the variable transformation
\be
x \E C_j \, g(t) \> , \quad C_j > 0 , \quad t \in [-\infty,+\infty]
\ee
gives the integral
\be
I_j \E \int_{-\infty}^{+\infty} dt \> C_j\, g'(t) \,  f \left ( C_j g(t) \right ) \> O_j \lrp C_j y g(y) \rrp
\> \simeq \>  
C_j h \sum_{k= -\infty}^{+\infty} 
g'(k h) \,  f \left ( C_j g(k h) \right ) \, O_j \left ( C_j y  g(k h) \right )
\label{trapez int}
\ee
and its trapezoidal approximation as usual. However, this time one requires 
\bea 
g(-\infty) \EA 0 \> , \quad  g'(-\infty) \E 0 
\label{requirm neg}\\
g(t) & \sim &  t \qquad {\rm for} \qquad t \to + \infty \> .
\label{requirm pos}
\eea
Then one has at large positive $ k $
\be
f \left ( C_j g(kh) \right ) \, O_j \left ( C_j y g(kh) \right ) \> \sim \>  f \left ( C_j k h \right ) 
 \, O_j \left ( C_j y k h  \right )
\E f \left ( C_j k h \right ) 
 \, \lrp \frac{ \sin \left ( C_j y k h /j \right )}{C_j y k h /j} \rrp^j \> .
\ee
If one chooses the free constant as
\be 
C_j \E \frac{j \pi}{h y} \> ,
\ee
i. e. such that at large $k$ the zeroes of the oscillating function $O_j$ are always taken, then
\be 
\sin \left ( C_j y k h /j \right ) \E \sin (k \pi) \E 0 \> .
\ee
This means that one can truncate the summation in Eq. (\ref{trapez int}) at some moderate positive $ k $ . 
For large
negative $ k $ the summation is restricted due to constraints in Eq. (\ref{requirm neg}).
Ooura and Mori have given a function $g(t)$ which satisfies all these requirements, viz.
\be
\boxed{
g_{\rm OM}(t) \E \frac{t}{1 - \exp(-2 \lambda \sinh t)} \> } \quad, \quad \lambda > 0 \> , 
\quad g_{\rm OM}(0) \E \frac{1}{2 \lambda} \> , \quad g'_{\rm OM}(0) \E \frac{1}{2} \> .
\label{OoMori func}
\ee
Indeed for $ t \to \pm \infty $ function values and derivatives 
approach the required limits in the typical double exponential way
\bea
g_{\rm OM}(t)  & \to & t \, \Theta(t) + |t| \exp \left [ -\lambda e^{|t|} \right ] \non
g'_{\rm OM}(t)  & \to & \Theta(t) - \lambda t  e^{|t|} \, \exp \left [ -\lambda e^{|t|}\right ] \> .
\label{asy g(t)}
\eea
\vspace{0.5cm}

\noindent
Thus 
\be
\boxed{
I_j[f] \E \int_0^{\infty} dx \> f(x) \, O_j(xy)  \> \simeq \> \frac{j \pi}{y} \, 
\sum_{k=-k_{\rm max}}^{+k_{\rm max}}
\, w_k \, f \lrp \frac{j \xi_k}{y} \rrp \> \lrp \frac{\sin \xi_k}{\xi_k} \rrp^j
\label{Ij DE}
}
\ee
with 
\be
w_k  \E g'_{\rm OM} (k h) \quad, \qquad  \xi_k \E  \frac{\pi}{h} \, g_{\rm OM} (k h) \> .
\label{wk xik}
\ee
Note that the weights $w_k$ and the abscissas $\xi_k$ in Eq. \eqref{wk xik} are independent of $ j$ and $ y $ . \footnote{The form \eqref{Ij DE} corresponds to the original integral after the substitution $ x = j \xi/y $.}

\vspace{1cm}

\noindent
There are several advantages of the OM method:
\bdes
\item[a)] There is no need to cut off the infinite integral at a large value $ \> x = b \> $ and
add the asymptotic contribution. Of course, there is an implicit cut-off for the summation over 
$ \> |k| < k_{\rm max} \> $ which turns into a choice of the stepsize for the trapezoidal integration.
From the asymptotic behavior in Eq. \eqref{asy g(t)} we choose it as
\be
\exp \lsp - \lambda \, \exp\lrp k_{\rm max} h \rrp \rsp \> \le \>  \epsilon \quad \Longrightarrow \> \> h \E  \frac{1}{k_{\rm max}} \ln  \lsp -\frac{1}{\lambda} \ln (\epsilon) \rsp
\label{stepsize}
\ee
so that the weights are sufficiently small at $ \> k = - k_{\rm max} \> $ and sufficiently close to $ 1 $ 
at $ \> k = +k_{\rm max} \> $.
Typically we take  
\be 
\epsilon \E  10^{-12} \q,\q \> \lambda \E 1/2 \> .
\label{accuracy}
\ee
\item[b)] Abscissas and weights for different values of the external parameter $ \> y \> $ are easily calculated 
by a simple division or by an overall rescaling (see Eqs. \eqref{wk xik} and \eqref{Ij DE}).
\item[c)] Only elementary functions are needed.
\item[d)] Automatic programs in Fortran and C are already available and can be downloaded from\\
\hspace*{2cm}  http://www.kurims.kyoto-u.ac.jp/~ooura/intde.html.\\
In particular, for the present case the routine \\
\hspace*{2cm} {\it intdeo : integrator of f(x) over (a,infinity),
\quad f(x) is oscillatory function} \\can be used.
\edes
\vspace{0.3cm}

\vspace{0.5cm}

\begin{table}[htb]
\begin{center}
\begin{tabular}{|r|c|c|c|c|} \hline
              &                      &                       &                      &                    \\
 $y \quad$   &  $ \> I_1[f_0] \> $  &  $ \> I_1 [f_1] \> $  &  $ \> I_2[f_0] \> $  & $ \> I_2[f_1] \> $  \\
              &                      &                       &                      &              \\ \hline
              &                      &                       &                      &                  \\
$0.01  $      &  \q $-3.04 \, (-08) $ \q   &  $ || < \,(-15)$         & $-6.07 \, (-08) $      &  $-1.68 \, (-13) $  \\
$0.02  $      &  \q $-1.52 \, (-08) $ \q   &  $ || < \,(-15)$         & $-3.04 \, (-08) $      &  $-2.81 \, (-14) $  \\
$0.05  $      &  \q $-6.08 \, (-09) $ \q   &  $ || < \,(-15)$         & $-1.21 \, (-08) $      &  $ \> \> \>  1.83 \, (-14) $  \\
$0.1 \>\>$    & \q  $-3.04 \, (-08) $ \q   &  $ || < \, (-15)   $     & $-6.08 \, (-09) $      &  $||  < \, (-15) $  \\
$0.2 \>\> $   &  \q $-1.54 \, (-09) $ \q   &  $ || < \, (-15)   $     & $-3.06 \, (-09) $      &  $ -2.04 \, (-15) $  \\
$0.5 \>\> $   &  \q $-6.54 \, (-10) $ \q   &  $ || < \, (-15)   $     & $-1.26 \, (-09) $      &  $ || < \, (-15) $  \\
$1 \q $       & \q  $-3.86 \, (-09) $ \q   &  $ || < \, (-15)   $     & $-6.92 \, (-10) $      &  $ || < \, (-15) $  \\
$2 \q $       &  \q $-2.74 \, (-10) $ \q   &  $ || < \, (-15)   $     & $-4.31 \, (-10) $      &  $ || < \, (-15) $  \\
$5 \q$        &  \q $-2.21 \, (-10) $ \q   &  $ || < \, (-15)   $     & $-2.90 \, (-10) $      &  $ || < \, (-15) $  \\
$10 \q \,$    &  \q $-2.06 \, (-10) $ \q   &  $ || < \, (-15)   $     & $-2.45 \, (-10) $      &  $ || < \, (-15) $  \\
$20  \q \,$   & \q  $-2.00 \, (-10) $ \q   &  $ || < \, (-15)   $     & $-2.21 \, (-10) $      &  $ \> \>  \> 4.28 \, (-15)$ \\
$50  \q \,$   & \q  $-1.96 \, (-10) $ \q   &  $ 4.19\, (-14) $     & $-3.38 \, (-10) $      &  $ -4.82 \, (-08)$ \\
$100  \q \>$  &  \q $-1.94 \, (-10) $ \q   &  $4.79 \, (-12) $     & $-2.92 \, (-07) $      &  $-6.90\, (-05) $ \\
$200  \q \> $ & \q  $-1.94 \, (-10) $ \q   &  $4.68 \, (-11) $     & ${\bf -1.10 \, (-05)} $ &  ${\bf -2.40 \, (-03)} $ \\
$500  \q \> $ &  \q $-1.93 \, (-10) $ \q   &  $1.83 \, (-10) $     & ${\bf -1.17 \, (-04)} $ &  ${\bf -2.54 \, (-02)} $ \\
$1,000 \q \> $ &  \q $-1.93 \, (-10) $ \q   &  $2.88 \, (-10) $     & ${\bf -2.93 \, (-04)} $ &  ${\bf -6.57 \, (-02)} $ \\
$2,000 \q \>$  &  \q $-1.93 \, (-10) $ \q   &  $3.62 \, (-10) $     & ${\bf -4.97 \, (-04)} $ &  ${\bf -1.19 \, (-01)} $ \\
$5,000 \q \> $ &  \q $-1.93 \, (-10) $ \q   &  $4.14 \, (-10) $     & ${\bf -7.29 \, (-04)} $ &  ${\bf -1.95 \, (-01)} $ \\
$10,000 \q \> $&  \q $-1.93 \, (-10) $ \q  &  $4.34 \, (-10) $     & ${\bf -8.52 \, (-04)} $ &  ${\bf -2.49 \, (-01)} $ \\ 
$20,000 \q \> $ &  \q $-1.93 \, (-10) $ \q &  $4.44 \, (-10) $     & ${\bf -9.34 \, (-04)} $ &  ${\bf -2.99 \, (-01)} $ \\ 
$50,000 \q \> $ &  \q $-1.93 \, (-10) $ \q &  $4.49 \, (-10) $     & ${\bf -9.98 \, (-04)} $ &  ${\bf -3.57 \, (-01)} $ \\
$100,000 \q \> \> $ &  \q $-1.93 \, (-10) $ \q &  $4.51 \, (-10) $     & ${\bf -1.02 \, (-03)} $ &  ${\bf -3.95 \, (-01)} $ \\
              &                            &                       &                       &       \\ \hline
  
\end{tabular}

\end{center}
\caption{Relative error for the different oscillatory integrals obtained with 
the double exponential method of Ooura \& Mori using $ \> 2 k_{\rm max}+ 1 = 289 \> $ function calls. Throughout the table
the abbreviation $ 3.0 \, (-08) $ for $ 3.0 \cdot 10^{-8} $ etc. is employed and the parameters of Eq. \eqref{accuracy}
are used. $ ||  < \, (-15)  $ indicates that
the absolute value of the relative deviation is smaller than $10^{-15}$ and thus subject to rounding errors in
double precision arithmetic.
Errors {\bf larger in magnitude} than those obtained by the Filon-Simpson method with $ N = 288 $ are printed in boldface.}
\label{tab:OM}
\end{table}
\vspace{0.5cm}


A disadvantage is that  $ \> y = 0 \> $ must be treated separately 
and does not reduce automatically to the standard
tanh-sinh method of Takashi and Mori for the integral over the non-oscillatory function $ \> f(x) \> $.
Also -- in contrast to the Filon-Simpson method -- the correct asymptotic behaviour of the oscillatory integrals
for $ \> y \to \infty \> $ -- is not built in. 
This shows up in the numerical results for the relative error
\be 
\epsilon_{\rm rel} \Def \frac{I_j^{\rm OM}[f_l] - I_j^{\rm exact}[f_l]}{I_j^{\rm exact}[f_l]} \q , \q  j = 1, 2 \q l = 0, 1
\ee
for small and large $ y $ collected in Table \ref{tab:OM}: whereas the 
Ooura-Mori method yields superior results when applied to the test functions with the oscillatory weight
$ \> O_1 = \sin(x y)/(x y) \> $ it starts to deteriorate for the oscillatory weight 
$ \> O_2 = 4 \sin^2(x y/2)/(x y)^2 \> $ when $ y $ becomes large so that finally the Filon-Simpson method 
takes the lead. Of course, this can be remedied by enlarging $ \> k_{\rm max} \> $ and/or taking a smaller stepsize
$ h $ as Table 2 demonstrates but this makes the method much less efficient in terms of function calls. 

Changing the value of the parameter $ \lambda $ in the OM method is of no help either: for example, 
the relative error of $ I_2^{\rm OM}[f_1] $ for $ \> y = 100,000 \> $ (last item in the last line of Table 1) becomes 
$ - 0.431 $ when $ \lambda = 3 $ is taken as in Ref. \cite{OoMori} \footnote{Ref. \cite{OMrobust} 
notes ``the magnitude of ($\lambda$) does not significantly affect the efficiency of the formula'' and takes
$ \lambda = \pi$ which subsequently is also used in Ref. \cite{BBBZ}.}.
\vspace{1.5cm}

\begin{table}[htb]
\begin{center}
\begin{tabular}{|r|cc|cc|} \hline
                  &                                  &                               &                           &   \\
 $y \Q $          & \hspace*{0.9cm} $ I_2[f_0] \> , \Q |\epsilon_{\rm rel}| < 10^{-6} $  & & \hspace*{0.9cm} $ I_2 [f_1] \> , \Q |\epsilon_{\rm rel}|< 10^{-3} $ &  \\
                  &                                  &                             &                             &                        \\
                  &  OM  \hspace*{1cm}               & FS \hspace*{1cm}            &   OM  \hspace*{1cm}         & FS \hspace*{1cm}      \\ \hline
                  &                                  &                             &                             &                       \\
\q $100  \q \>$   &   $ 249 $    \hspace*{1cm}       &  $632$   \hspace*{1cm}      &  $   189 $   \hspace*{1cm}     & $ 308 $  \hspace*{1cm}    \\
\q $200  \q \> $  &   $ 295 $    \hspace*{1cm}       &  $674$   \hspace*{1cm}      &  $ {\bf  353} $  \hspace*{1cm} &   $ 350 $  \hspace*{1cm}    \\
\q $500  \q \> $  &   $ {\bf 907} $  \hspace*{1cm}    & $594$  \hspace*{1cm}       &  $ {\bf  817} $  \hspace*{1cm} &   $ 394 $  \hspace*{1cm}  \\
\q $1,000 \q \> $ &   $ {\bf 1,565} $ \hspace*{1.3cm}   & $498$   \hspace*{1cm}      &  $ {\bf 1,553} $  \hspace*{1.2cm}  &   $ 418 $  \hspace*{1cm}  \\
\q $2,000 \q \>$  &   $ {\bf 2,673} $  \hspace*{1.3cm}  & $400$  \hspace*{1cm}       &  $ {\bf 2,965} $ \hspace*{1.2cm}  &   $ 438 $  \hspace*{1cm}  \\
\q $5,000 \q \> $ &   $ {\bf 5,315} $  \hspace*{1.3cm}  & $288$ \hspace*{1cm}        &  $ {\bf 7,010} $  \hspace*{1.2cm} &   $ 458 $  \hspace*{1cm}  \\
\q $10,000 \q \> $&   $ {\bf 8,770} $  \hspace*{1.3cm}  & $220$ \hspace*{1cm}        &  $ {\bf 13,490} $ \hspace*{1.4cm} &   $ 474 $  \hspace*{1cm}  \\ 
\q $20,000 \q \> $ &  $ {\bf 14,230} $ \hspace*{1.5cm}  & $166$ \hspace*{1cm}        &  $ {\bf 26,010} $ \hspace*{1.4cm} &   $ 484 $  \hspace*{1cm}  \\ 
\q $50,000 \q \> $ &  $ {\bf 26,000} $ \hspace*{1.5cm}  & $112$  \hspace*{1cm}       &  $ {\bf 62,200} $ \hspace*{1.4cm} &   $ 496 $  \hspace*{1cm}  \\
\q $100,000 \q \> \, $ &  $ {\bf 39,740} $ \hspace*{1.5cm} & $\> 82$  \hspace*{1cm}  &  $ {\bf 120,500} \> $\hspace*{1.6cm} & $ 504 $  \hspace*{1cm}  \\
                  &                                   &                            &                                 &                   \\ \hline 
                      
\end{tabular}

\end{center}
\caption{Number of function calls required for obtaining a magnitude of the relative error less than $ 10^{-6} $ for $ \> I_2[f_0] \> $ and $ 10^{-3} $ for $ \> I_2 [f_1] \>  $, respectively, with the Ooura-Mori (OM) and the Filon-Simpson (FS) method for oscillatory integrals.
As in Table~1 the parameters are those of Eq. \eqref{accuracy} and cases where the OM method is inferior to the FS method are printed
in boldface.}
\label{tab:N}
\end{table}
\vspace{1cm}
 
How can one understand the results of the OM method for our test integrals at large $ y $ ? \\
\noindent
This is straightforward for functions $ f_0(x) $ which do not vanish at the origin (in our test example the function $f_0(x) = e^{-x} $ ) 
because
\be 
I_j^{\rm OM}[f_0] \E \frac{j \pi}{y} \, \sum_{k=-k_{\rm max}}^{+k_{\rm max}}
\, w_k \, f_0 \lrp \frac{j \xi_k}{y} \rrp \> \lrp \frac{\sin \xi_k}{\xi_k} \rrp^j \yto \, f_0(0) \, \frac{j \pi}{y} \, 
\sum_{k=-k_{\rm max}}^{+k_{\rm max}} \, w_k \lrp \frac{\sin \xi_k}{\xi_k} \rrp^j
\ee
whereas the exact asymptotic value from Eqs. \eqref{I1 asy} and \eqref{I2 asy} is
\be
I_j^{\rm exact}[f_0] \yto \frac{j \pi}{2 y} \, f_0(0) \> .
\ee
Therefore
\be 
\frac{I_j^{\rm OM}[f_0] - I_j^{\rm exact}[f_0]}{I_j^{\rm exact}[f_0]} \yto  2 \sum_{k=-k_{\rm max}}^{+k_{\rm max}}
\, w_k \, \lrp \frac{\sin \xi_k}{\xi_k} \rrp^j - 1 \deF \delta_j^{(0)}
\ee
where the "defects" $ \delta_j^{(0)} $ are universal, i.e. do neither depend on $ y $ nor on $ f_0(0) $ but only on the step-size $h$
and the parameter $ \lambda $ (when $k_{\rm max} $ is large enough), see Eq. \eqref{wk xik}.
For the results presented in Table 1 these constants have the value
\be 
\delta_1^{(0)} \E - 1.926 \cdot 10^{-10} \Q, \Q
\delta_2^{(0)} \E - 1.059 \cdot 10^{-3}
\label{delta12 0}
\ee
That the relative error of $ \> \> I_j^{\rm OM} [f_0] \> , \>  \> j = 1, 2\>  $
approaches these constants is clearly seen in the results displayed in Table 1. It is
due to the well-known fact that in the large-$y$ limit the constant function $ f_0(0) $
is {\it not} integrated exactly in the double-exponential scheme. Neither are low-order polynomials and therefore
we have for functions $ f_1(x) $ where $ f_1(0) = 0 , \> f_1'(0) \neq 0 $ 
\be 
I_j^{\rm OM}[f_1] \yto  f'_1(0) \, \frac{j^2 \pi}{y^2} \, 
\sum_{k=-k_{\rm max}}^{+k_{\rm max}}
\, w_k \, \xi_k \, \lrp \frac{\sin \xi_k}{\xi_k} \rrp^j \> .
\label{Ij 1 OM}
\ee
Utilizing Eq. \eqref{I1 asy} with the upper limit $ b \to \infty $ we thus find that
for integrals with the oscillating factor $ O_1 $ the relative error of the OM method still approaches 
a constant defect
\be  
\frac{I_1^{\rm OM}[f_1] - I_1^{\rm exact}[f_1]}{I_1^{\rm exact}[f_1]} \yto  \pi \sum_{k=-k_{\rm max}}^{+k_{\rm max}}
\, w_k \, \sin \xi_k - 1 \deF \delta_1^{(1)} \> .
\ee
For the parameters used in Table 1 we find
\be 
\delta_1^{(1)} \E 4.548 \cdot 10^{-10}
\ee
which agrees well with the results of Table 1 at high $y$.

The situation is different for integrals with the oscillating factor $ O_2 $ and vanishing function value at $ x = 0 $
(in our test example the function $f_1(x) = x \, e^{-x} $ ): Eq. \eqref{I2 asy} then shows a {\it logarithmic}
enhancement of the exact integral at asymptotic large values of  $ y $   
\be 
I_2[f_1] \yto \frac{2}{y^2} \lsp f_1'(0) \ln y + C(\infty) \rsp \> .
\ee
In contrast the OM approximation in Eq \eqref{Ij 1 OM} cannot develop a logarithmic dependence for 
finite $ k_{\rm max} $. Thus
\bea  
\frac{I_2^{\rm OM}[f_1] - I_1^{\rm exact}[f_1]}{I_1^{\rm exact}[f_1]} &\yto&  \frac{2 \pi}{\ln y + 
C(\infty)/f'(0)} \sum_{k=-k_{\rm max}}^{+k_{\rm max}}
\, w_k \, \frac{\sin^2 \xi_k}{\xi_k} - 1 \non 
&\deF& \delta_2^{(1)}/\ln y - 1 + {\cal O} \lrp \frac{1}{\ln^2 y} \rrp \> .
\label{I21 log dep}
\eea
For the parameters of Table 1 we find
\be 
\delta_2^{(1)} \E 6.9735
\ee
and thus from Eq. \eqref{I21 log dep} the predictions $ \>  I_2[f1]( y = 100,000 ) = -0.3943 \> $ 
and $\> I_2[f1]( y = 50,000 ) = -0.3555 \> $
which are in good agreement \footnote{Note that 
$ C(\infty) = 0 $ for the function $ f_1(x) = x e^{-x} $.} with the last two entries of Table 1.
Since the logarithmic enhancement of the exact integral always overwhelms the power-like behaviour of the 
OM approximation the relative deviation therefore will approach the value $ - 1 $ asymptotically.

Obviously this breakdown of the OM method is due to the fact that the weight function 
$ \> O_2 \> $ is always positive or zero, i.e. does not really oscillate. This then gives rise to 
logarithmic terms in the exact integrals (see Eq. \eqref{exact2}).
Note that the Filon-Simpson method has built in these logarithmic terms
as can be seen from Eqs. \eqref{log terms 1}, \eqref{log terms 2} and therefore copes much better with the limit 
$ y \to \infty $. This is clearly demonstrated in Table 2.

\vspace{2.5cm}

\section{Summary}

Relatively simple and straightforward quadrature rules of
Filon-Simpson form have been presented which are applicable for numerical 
integration 
of oscillatory integrals of the type (\ref{I1}, \ref{I2}). They employ 
equidistant integration points (including the endpoints) and 
weights which have to be calculated anew for each value of the frequency parameter 
$y$. The choice of equidistant points allows easy construction of extended
Filon-Simpson quadrature rules so that the accuracy of the result can be simply
assessed by increasing the number of subdivisions. 
Inevitably the Filon-Simpson weights are more involved 
than the ones of standard quadrature rules as they are given in terms
of sine and cosine integrals and elementary functions. However, the price 
for an accurate evaluation of these weights is modest and worthwhile as the
Filon-Simpson quadrature rules not only reduce to the ordinary Simpson 
rule for $y = 0$ but also give the leading and subleading terms for $y \to \infty$ 
provided smooth functions are integrated and the spacing of integration points 
is fine enough. Although a rigorous error estimate has
yet to be given, numerical tests have shown that in this regime they
do far better than standard quadrature rules. A detailed numerical comparison is made
with the double-exponential method proposed by Ooura and Mori: whereas this method 
gives superior results for Fourier-sine integrals it requires much more function
calls than the Filon-Simpson method when applied to $ \sin^2 (x y) $-type integrands 
with large values of the frequency parameter $ y $.
\vspace{0.1cm}

Given its built-in properties for small and large $ y $
the Filon-Simpson method is thus an attractive option for all applications
where these types of oscillating integrals have to be evaluated. 

\vspace{2.5cm}

\noindent
{\bf Acknowledgement:} I would like to thank Prof. A. Iserles for a very kind and helpful
correspondence. \\
{\it Habent sua fata libelli:} The first version of this paper was sent to J. Comput. Phys. in 2006
where one referee found it ``well written and a nice contribution",
whereas the second one didn't see ``sufficient meat'' and 
urged me to write a totally different (``worthwhile'') paper following his ideas.
Although the editor, Prof. Lang, tried to find other solutions this was unacceptable for me. 
In the end the paper lay dormant until I learned about the ingenious method of Ooura and Mori. 
This happened when I was waiting at a printer, glancing at some of the printouts and
rekindled my interest in efficient computation of oscillatory integrals. 
I am indebted to the unknown colleague who printed out Ref. \cite{OoMori}
just in the right moment...

\vspace{2.5cm}

\bce
{\Large\bf Appendix}
\ece

\renewcommand{\thesection}{\Alph{section}}
\renewcommand{\theequation}{\thesection.\arabic{equation}}
\setcounter{equation}{0}
\setcounter{section}{1}

\vspace{0.2cm}

\subsection{Exact asymptotic behaviour of the integrals}

Here we derive the asymptotic expansion of the oscillatory integrals beyond
the leading order which was only considered in the main text. We assume that the
function $f(x)$ is analytic at $ x = 0 $ and that the upper limit $b$ is finite.

We start with the integral $I_1$ whose asymptotic expansion is easy
to obtain by a subtraction followed by an integration by parts so that an 
additional inverse power of $y$ is generated:
\bea
I_1[f](a=0,b,y) \EA \int_0^b dx \> f(0) \, \frac{\sin xy}{xy} +  \frac{1}{y} 
\int_0^b dx \> \frac{f(x) - f(0)}{x} \, \sin xy \E \frac{f(0)}{y} F_0^{(1)}(by) 
\non 
&& + \frac{1}{y}  \left [ - \frac{f(x) - f(0)}{x}  
\frac{\cos xy}{y} \, \Biggr|_0^b + \frac{1}{y} \int_0^b dx \, 
\left (\frac{f(x) - f(0)}{x} \right )' \cos xy \, \right ] .
\eea
Repeating the process in the remaining integral it is seen that it is of higher 
order. By using the explicit expression for  $F_0^{(1)}(by)$ given in Eq. 
(\ref{F^1}) together with the asymptotic expansion (\ref{sinint asy}) we thus 
obtain
\be
I_1[f](a=0,b,y)  \> \yto \> 
\frac{\pi f(0)}{2y} + \frac{1}{y^2} \, \left [ f'(0) - f(b) \frac{\cos yb}{b} 
\right ] + {\cal O} \left ( \frac{\sin by}{y^3} \right ) \> .
\label{I1 asy}
\ee
Similarly we obtain for the integral $I_2[f](a=0,b,y)$ after {\it two} subtractions
\be
I_2[f](a=0,b,y) \E   \frac{f(0)}{y} F_0^{(2)}(by) + 
\frac{f'(0)}{y^2}  F_1^{(2)}(by)
+ \frac{2}{y^2} \int_0^b dx \> g(x) \left ( 1 - \cos xy \right ) 
\ee
where $g(x) = (f(x) - f(0) - x f'(0))/x^2 $ is regular at $x = 0 $. Therefore we 
may apply the procedure of repeated integration by parts to obtain for the last 
term
\be 
\frac{2}{y^2} \int_0^b dx \> g(x) \left ( 1 - \cos xy \right ) \E 
\frac{2}{y^2} \int_0^b dx \> g(x) - \frac{2}{y^3} \, g(x) \sin xy  \Biggr|_0^b  
+ \ldots \> .
\ee
Finally by employing the explicit expressions for $F_j^{(2)}(by)$ given in 
Eq. (\ref{F^2}) together with the asymptotic expansions (\ref{sinint asy}) we 
obtain
\be
I_2[f](a=0,b,y) \> \yto \> \frac{\pi f(0)}{y}
+ \frac{2}{y^2} \left [ \, f'(0) \, \ln y + C(b) \, \right ] - 
\frac{2 f(b) \sin by}{b^2 y^3} + \ldots \> .
\label{I2 asy}
\ee
Formally the last term in Eq. (\ref{I2 asy}) is of next-to-next-to-leading order
but it is needed to verify the relation (\ref{I1 from I2}) 
to next-to-leading order.
Note also the appearance of logarithmic terms in the asymptotic expansion of 
$I_2 $ ; the constant $C(b)$ in Eq. (\ref{I2 asy}) is given by
\be
C(b) \E \left ( \, \gamma + \ln b \, \right ) \, f'(0) - \frac{f(0)}{b} + 
\int_0^b dx \> \frac{f(x) - f(0) - 
x f'(0)}{x^2}
\> .
\label{C(b) 0}
\ee
Three integration by parts in the last integral bring it into the form
\be
C(b) \E \gamma f'(0) - \frac{f(b)}{b} + f'(b) + \left [ \, f'(b) - b f''(b)
\, \right ]\, \ln b + \int_0^b dx \> x \ln x \, f'''(x) 
\label{C(b) 1}
\ee
which will be needed for comparison with the extended Filon-Simpson rule.

It is clear that nothing prevents the procedure to be extended to arbitrary order
but for our purposes this is not needed.  
One may check the asymptotic expansions (\ref{I1 asy}, \ref{I2 asy}) by applying
them to the functions $f_l(x)$ used for the numerical tests: in the limit
$b \to \infty$ one obtains full agreement with the first terms of the asymptotic 
expansion of the exact results (\ref{exact1}, \ref{exact2}). 

\vspace{0.5cm}

\subsection{Asymptotic behaviour of the Filon-Simpson rules}

How do the Filon-Simpson quadrature rules behave in the asymptotic limit ? To 
answer this question we just have to plug the asymptotic expansions of the 
functions $ F_i^{(j)}(by) $ into the expressions for the weights. After 
some algebraic work we obtain
\be
\sum_{i=0}^2 w_i^{(1)} \, f_i \E \frac{1}{y} \frac{\pi}{2} f_0 + \frac{1}{y^2} 
\left [ \, \rightdelta f_0 - \frac{f_2 \cos by}{b} \, \right ] 
+ {\cal O} \left ( \frac{\sin by}{y^3} \right ) \> , \hspace{0.5cm} a \E 0, \> 
b = 2 h
\label{FS1 asy 1}
\ee
where $f_i \equiv f(ih) $ and
\be
\rightdelta f_0 \E \frac{1}{2 h} \, \left [ \, - 3 f(0) + 4 f(h) - f(2h) \, 
\right ] \> \hto \> f'(0) - \frac{1}{3} h^2 f'''(0) + \ldots \> .
\label{def delta1}
\ee
Thus one obtains the correct subleading term of the asymptotic expansion
except that the derivative of the function at $x = 0$ is replaced by its (forward)
finite-difference approximation $\rightdelta f_0$.

Similarly, one finds that 
for the integral $I_2[f]$ the Filon-Simpson quadrature rule has the asymptotic 
expansion
\be
\sum_{i=0}^2 w_i^{(2)} \, f_i \E \frac{1}{y} \pi f_0 + \frac{2}{y^2} \left [ 
\, \rightdelta f_0  \cdot \ln y  + C_{FS}(b)  \right ] + 
{\cal O} \left ( \frac{\sin by}{y^3} \right ) \hspace{0.5cm} a \E 0, \> b = 2 h
\label{FS2 asy 1}
\ee
where the constant is given by
\be
C_{FS}(b) \E \left ( \, \gamma + \ln b \, \right ) \, 
\rightdelta f_0 - \frac{f_0}{b} + \frac{b}{2} \, \delta^2 f_1 \> .
\ee
Here 
\be
\delta^2 f_1 \E \frac{1}{h^2} \, \left [ \, f(0)  -2 f(h) + f(2h) \, \right ] \> 
\hto \> f''(0) + h f'''(0) + \ldots = f''(h) + {\cal O} \left ( h^2 \right )
\label{def delta2}
\ee
is a finite-difference approximation to the second derivative of the function 
$f(x)$ at $x = 0 $ (or better at $x = h$).
Comparing with the exact next-to-leading term in the asymptotic expansion 
of $I_2[f]$ we see that again  finite differences are subsituted for
derivatives and that
the last integral in Eq. (\ref{C(b) 0}) is replaced by
\be
\int_0^b dx \> \frac{f(x) - f(0) - x f'(0)}{x^2} \> \simeq \>  
\int_0^b dx \> \left [ \, \frac{1}{2} f''(0) + \ldots \, \right ] \E 
\frac{b}{2}  f''(0) + \ldots 
\ee
which is valid for regular functions and increments $b = 2h$ which are small 
enough.

\vspace{0.5cm}

\subsection{Asymptotic behaviour of the extended Filon-Simpson rules}

Let us now investigate how the extended Filon-Simpson rules behave in the limit
$y \to \infty$. To do that we need the asymptotic behaviour of the simple rules
in intervals with non-zero lower limit. Using the quadrature rules in the form 
\be
\sum_{i=0}^2 w_i^{(j)} f_i \E \left [ \, f_0 - a \rightdelta f_0 + 
\frac{a^2}{2} \delta^2 f_1 \, \right ] \, J_0^{(j)} +  \left [ \, \rightdelta f_0 - 
a \, \delta^2 f_1 \, \right ] \, 
J_1^{(j)} + \frac{1}{2} \delta^2 f_1 \, J_2^{(j)} 
\label{FS sum 2}
\ee
we obtain for the Filon-Simpson quadrature of $I_1[f]$
\be
\sum_{i=0}^2 w_i^{(1)} f_i \> \yto \> 
\frac{1}{y^2} \, \left [ \, \frac{f_0}{a} \cos ay -  \frac{f_2}{b} \cos by \, 
\right ] \> , \hspace{0.4cm} a \ne 0, \> b =  a + 2 h
\ee
and therefore
\bea
\sum_{i=0}^N w_i^{(1)} f_i & \yto & \frac{\pi f_0}{2 y} + \frac{1}{y^2} \, 
\left [ \, \rightdelta f_0 - f_2 \frac{\cos (2hy)}{2h}
\, \right ] \non
&& \hspace{0.8cm} + \frac{1}{y^2} \, \left [ \, \frac{f_2}{2h} \cos (2hy) -  
\frac{f_4}{4h} \cos 4hy \, \right ] + \ldots \non
&& \hspace{0.1cm} \ldots \, + \frac{1}{y^2} \, \left [ \, \frac{f_{2N-2}}{(2N-2)h} 
\cos ((2N-2)hy) -  \frac{f_N}{Nh} \cos (2Nh) \, \right ]  \> .
\eea
Here the first line gives the contribution from the first interval $[0,2h]$
(see Eq. (\ref{FS1 asy 1})), the second
line the one from the second interval $[2h,4h]$ and so on. It is seen that in the 
$1/y^2$-terms the contributions cancel pairwise and only the ones from
the first and the last interval survive. Therefore
\be
\sum_{i=0}^N w_i^{(1)} f_i \> \yto \> 
\frac{\pi f_0}{2 y} + \frac{1}{y^2} \, \left [ \, \rightdelta f_0 - \frac{f_N}{Nh} 
\cos 2Nh \, \right ] + \ldots
\ee
which is again the correct asymptotic result (\ref{I1 asy}) except 
that the derivative of the function at $x = 0$ is replaced by the finite difference
$\rightdelta f_0$.

\vspace{0.2cm}
The subleading asymptotic terms for the Filon-Simpson quadrature of $I_2[f]$ are
more involved because of the logarithmic terms. From Eq. (\ref{FS sum 2}) and the 
asymptotic behaviour of the integrals $J_k^{(2)}$ one gets after some algebra
\bea
\sum_{i=0}^2 w_i^{(2)} f_i & \yto & 
\frac{2}{y^2} \, \Biggl [ \, \frac{f_0}{a}  -  \frac{f_2}{b} + 2 h \delta^2 f_1 + 
\left (  \leftdelta f_2 - b \, \delta^2 f_1 \right ) \, \ln b \non
&& \hspace{3.5cm} - \left (  \rightdelta f_0 - a \, \delta^2 f_1 \right ) \, 
\ln a \, \Biggr ] \> ,
\hspace{0.4cm} a \ne 0, \> b =  a + 2 h \> .
\eea
Here 
\be
\leftdelta f_2 \E \frac{1}{2h} \left [ f_0 - 4 f_1 + 3 f_2  
\right ] \> \hto \> f_2' - \frac{h^2}{3} f'''_2 + \ldots
\ee
is the backward finite-difference approximation for the derivative of the function
$f(x)$ at the point $x_2 = b = a + 2 h$. It is obtained from the (forward) form 
(\ref{def delta1}) by the exchange $ a \leftrightarrow b $. 

\noindent
Together with the result (\ref{FS2 asy 1}) for the first interval we therefore have
\bea
\sum_{i=0}^N w_i^{(2)} f_i & \yto &
\frac{\pi f_0}{y} + \frac{2}{y^2} \, \Biggl \{  \, \rightdelta f_0 \, ( \gamma 
+ \ln y ) + \frac{f_2 - f_0}{2h} - \frac{f_N}{Nh} + h  \delta^2 f_1 + 
2h \sum_{i=2}^{N/2}  \delta^2 f_{2i-1} \non
&& + \left ( \rightdelta f_0 -  \rightdelta f_2 + 2h  \delta^2 f_3 \right ) \, 
\ln(2h) + \left ( \leftdelta f_N - Nh  \delta^2 f_{N-1} \right ) \, \ln (Nh) \non
&& + 2h \sum_{i=2}^{N/2-1} \left [ \,  \frac{\leftdelta f_{2i} - 
 \rightdelta f_{2i}}{2h}  + 2ih \frac{\delta^2 f_{2i+1} - \delta^2 f_{2i-1}}{2h} 
  \, \right ] \, \ln (2ih) \, \Biggl \}  \> .
\label{exFS2 asy}
\eea
At first sight this looks rather complicated unless one recognizes the sums as 
(extended) trapezoidal rules with stepsize $2h$
for the corresponding integrals (see, e.g. Eq. (25.4.2) in Ref. \cite{Handbook})
\be
2h \, \left [ \, \frac{g_0}{2} + g_2 + g_4 + \ldots +
g_{N-2} + \frac{g_N}{2} \, \right ] \E 
\int_0^{Nh} dx \> g(x)  + {\cal O} \left ( h^3 \right ) \> .
\ee 
Furthermore
\bea
&& \frac{f_2 - f_0}{2h} \> \hto \> f_0' \> , \hspace{0.7cm} \rightdelta f_0 
- \rightdelta f_2 + 2h  \delta^2 f_3 \E {\cal O} \left ( h^2 \right) \> , \non
&& \frac{\leftdelta f_{2i} - \rightdelta f_{2i}}{2h} \E  {\cal O} \left ( h^3 
\right) \> , \hspace{0.7cm}  
\frac{\delta^2 f_{2i+1} - \delta^2 f_{2i-1}}{2h} \> \hto \> f'''_{2i} \> .
\eea
Therefore in the limit $ h \to 0 $ Eq. (\ref{exFS2 asy}) becomes
\bea
\sum_{i=0}^N w_i^{(2)} f_i & \yto &
\frac{\pi f_0}{y} + \frac{2}{y^2} \, \Biggl \{  \, f'(0) \, ( \gamma + \ln y )
+ f'(0) - \frac{f(b)}{b} + \int_0^b dx \> f''(x) \non
&& \hspace{0.8cm} + \left [ \, f'(b) - b f''(b) 
\, \right ] \, \ln b \, + \int_0^b dx \> x \ln x \, f'''(x) \Biggr \}
\eea
which agrees exactly with the subleading term of Eq. (\ref{I2 asy}) and the
form (\ref{C(b) 1}) of the constant $C(b)$.

\newpage

\end{document}